# Harnessing ultrafast optical pulses for 3D microfabrication by selective tweezing and immobilization of colloidal particles in an integrated system

*Krishangi Krishna, Jieliyue Sun, Wenyu Liu, Robert H. Hurt, Kimani C. Toussaint Jr.*\*

Krishangi Krishna, Jieliyue Sun, Wenyu Liu, Kimani C. Toussaint Jr.
PROBE Lab, School of Engineering, Brown University, Providence, RI 02912, USA.
E-mail: kimani_toussaint@brown.edu

Kimani C. Toussaint Jr.
Brown University Center for Digital Health, Providence, RI 02903, USA

Robert H. Hurt
Laboratory for Environmental and Health Nanoscience, School of Engineering, Providence, RI 02912, USA.

Microfabrication using nano- to micron-sized building blocks holds great potential for applications in next-generation electronics, optoelectronics, and advanced materials. However, traditional methods like chemical vapor deposition and molecular beam epitaxy require highly controlled environments and specialized equipment, limiting scalability and precision. To address these challenges, we present a single-laser platform for selective tweezing and immobilization of colloids (STIC) that integrates particle manipulation, assembly, and stabilization in one system. STIC utilizes a femtosecond laser at ultra-low power for precise, contact-free optical manipulation of colloids without material damage. At higher power, the same laser enables two-photon polymerization (TPP) to immobilize colloids securely in their intended positions. Using STIC, we demonstrate the assembly of 3D structures from dielectric beads to patterned arrangements of transition metal dichalcogenides (TMD e.g., MoS$_2$). We also incorporate a TPP-fabricated handle as an intermediate support, significantly enhancing the optical tweezing efficiency of TMDs. The single-laser design eliminates the need for dual-laser systems, simplifying optical alignment, reducing heating damage, and improving efficiency. Additionally, we show that STIC supports direct multiphoton imaging for *in situ* inspection during fabrication. This work establishes a versatile, scalable optical platform for high-precision microstructure fabrication, offering a pathway to overcome current limitations in micro- and nanomanufacturing.

## 1. Introduction

The ability to assemble arbitrary 3D structures using nano- to micron-sized building blocks opens a range of exciting applications. This approach facilitates the miniaturization of devices, paving the way for the construction of high-performance microelectromechanical systems (MEMS),[1] lab-on-a-chip platforms,[2] as well as novel photonic waveguides,[3] and 2D materials,[4] all of which rely on precision and control at the micron scale.

Over the past two decades, significant efforts have been dedicated to developing various synthesis strategies for high-quality micro and nano-fabricated structures. Techniques such as chemical vapor deposition (CVD),[5] electron beam lithography (EL),[6] and molecular beam epitaxy (MBE)[7] have been at the forefront of these advancements. These methods hold the promise of producing atomically thin layers of materials like graphene, transition metal dichalcogenides such as molybdenum sulfide ($MoS_2$) and hexagonal boron nitride (hBN), that are crucial for a range of applications including electronics, optoelectronics, and energy storage.[8–10] Despite the progress made, the fabrication of these 2D materials remains a challenging endeavor. Often requiring highly controlled environments such as ultra-high pressures ($10^{-6}$ torr) and temperatures (500°C) and sophisticated equipment to prevent contamination, and ensure proper deposition and crystallization.[11–13]

An all-optical fabrication platform provides a promising alternative for the synthesis of such structures. Optical tweezers (OT), as a contact-free tool, enables on-demand manipulation and placement of colloids in 3D space with nanometer-level precision. Researchers have applied OT on 1-μm diameter dielectric particles, functionalized with biotin and avidin, to create cross-linked structures when brought into contact. While this approach has been used to fabricate the largest structure comprising ~440 particles, it is not feasible on other types of particles since it relies on functionalized spheres.[13] Alternatively, when particles are immersed in photoresists, the colloids can be immobilized *in situ* through photopolymerization. The automation of this platform for fabricating 3D structures is being explored through the integration of image processing techniques and holographic optical tweezers (HOTs) via a spatial light modulator generating multiple trap sites within the sample plane. Current methods have been successfully employed for assembling microgranular crystals, shape-complementary planar microstructures, and quasi-crystalline photonic heterostructures, among others.[14] However, these approaches generally require two laser systems: a continuous-wave (CW) laser for optical trapping and a femtosecond (fs) laser for two-photon polymerization (TPP).[15] The need for dual laser systems complicates optical alignment, and the typical high average optical power used in optical trapping risks altering or damaging the functional materials.

To address these challenges, we propose a single-laser approach that achieves selective tweezing and immobilization of colloids (STIC) for fabricating novel materials.[16] We have previously demonstrated that under specific conditions, femtosecond (fs) laser-assisted selective holding with ultra-low power (FLASH-UP) significantly outperforms conventional CW-OT by a five-fold improvement in trap stiffness at ultra-low average powers on the order of tens of microwatts.[17] This enables the direct manipulation of micron-sized dielectrics and cells without observing any deleterious effects. By leveraging the power-threshold difference between TPP and FLASH-UP (approximately 20-30 mW), STIC is realized using a single fs laser beam at the same wavelength with different power settings for tweezing and polymerization. We demonstrate this platform by fabricating 3D microstructures comprising of dielectric beads, and further demonstrate the patterned assembly of transition metal dichalcogenide (TMDs), such as $MoS_2$, and graphene stabilized with sodium dodecyl sulfate (SDS) colloidal nanosheets. This system enables direct multiphoton imaging post-assembly, showcasing the capability of this integrated optical platform for colloid placement, assembly, and inspection.

## 2. Results and Discussion
### 2.1. Systematic design of arbitrary structures with dielectric microspheres

As a proof-of-concept, we apply STIC on micron-sized diameter particles. The manual assembly of a two-dimensional (2D) 3x3 grid using microspheres of different sizes is illustrated in **Figure 1**. In this process, alternating 2-μm and 5-μm diameter dielectric microspheres are arranged with an average spacing of 6 μm, as indicated by the yellow arrows in the figure. Achieving this precise arrangement requires careful control of the OT system, particularly in maintaining the laser's average power at the threshold intensity of 25 mW. Maintaining this threshold is crucial because any deviation could cause the scattering force of the laser to become dominant over the gradient force, resulting in the displacement of the particles to a different plane, thus compromising the assembly process.

This method is also successfully employed to fabricate more complex structures, as demonstrated in **Figure 2**. Specifically, the process can be utilized to create arbitrary structures, such as forming the word "BROWN" using only 2 μm diameter silica microspheres. The hydrophilic nature of these silica microspheres plays a significant role in the fabrication process. Because of their affinity for water, the microspheres can be positioned closer together, allowing for more compact and precise assemblies. This close positioning is advantageous in constructing intricate patterns and structures, highlighting the versatility and precision of the technique.

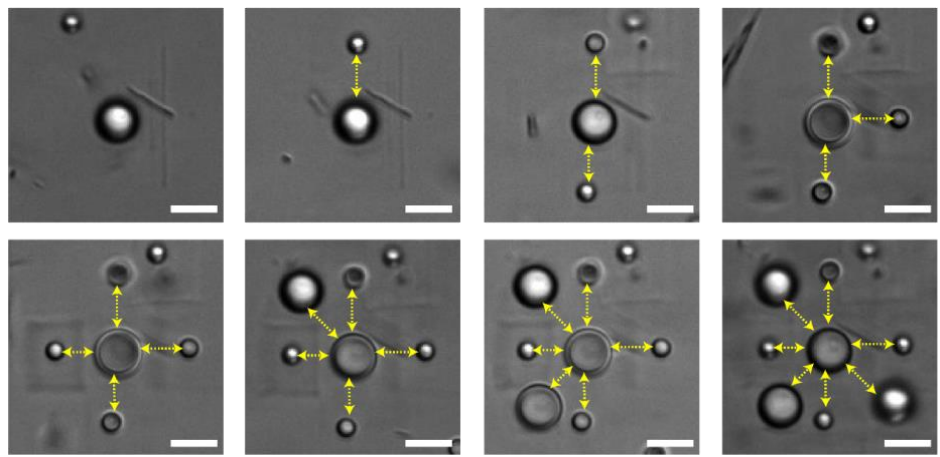

**Figure 1.** Time lapse sequence demonstrating 2D assembly with 2 μm and 5 μm dielectric microspheres. Yellow arrows demonstrate the spacing between particles. Scale bars, 5 μm.

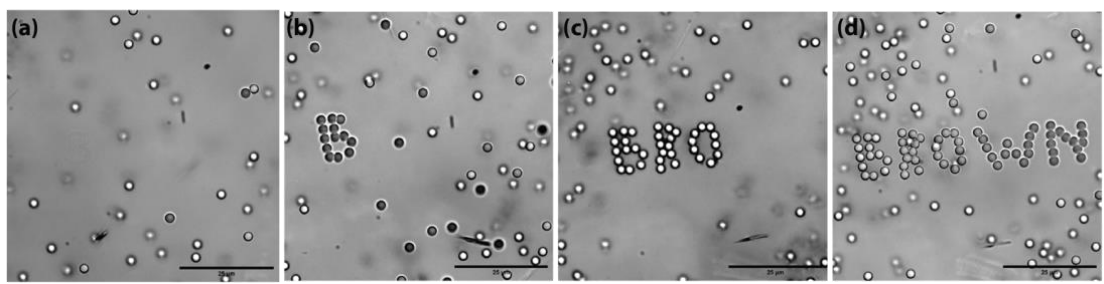

**Figure 2.** Time lapse sequence demonstrating 2D ability to use STIC to create arbitrary patterns in photoresist.

To fabricate a 2-layered 3D structure, we begin by assembling a 3x3 grid of 2 μm silica beads. Next, a 5 μm polystyrene bead is trapped and positioned 7 μm above this grid. The axial position of the polystyrene bead is carefully adjusted by lowering the objective lens until slight resistance between the polystyrene bead and the silica structure is observed due to the surface energy between hydrophobic and hydrophilic particles. At this point, the average laser power is increased, allowing the polystyrene bead to adhere to the silica grid. This adhesion is facilitated by polymerizing the hydrogel surrounding the particles, thereby securing the polystyrene bead in place and completing the 3D structure (**Figure 3**).

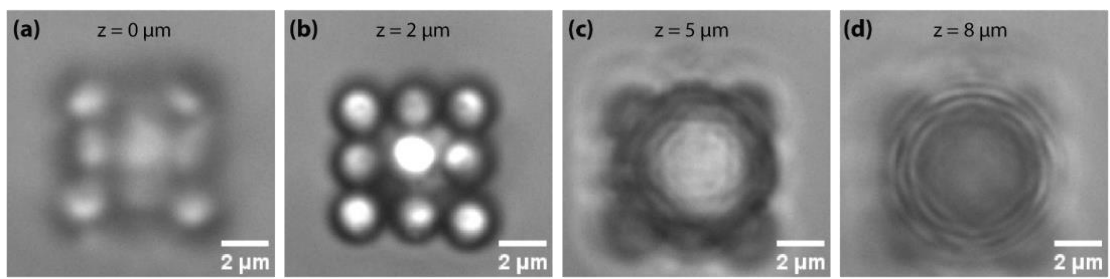

**Figure 3.** Time lapse sequence demonstrating 3D ability to create patterns with 2 μm and 5 μm dielectric microspheres. (**Video 1**).

We also successfully fabricated a 3-layered structure using 1 μm fluorescent Nile Red (535/575) particles, as shown in Supplemental **Figure S1**. This multilayered structure demonstrates the precision and versatility of our assembly technique. To accurately capture the two-photon fluorescent signal emitted by the Nile Red particles, we introduce a flip mirror into the optical setup. Positioned before the tube lens, this flip mirror redirects the two-photon fluorescent signal into the photomultiplier tube, ensuring efficient detection and measurement of the fluorescence (Video 2). This setup permits close monitoring of the assembly process and verifies the correct placement and orientation of the fluorescent particles within the structure.

## 2.2. Systematic design of arbitrary structures with graphene

Next, we apply STIC to the manipulation of surfactant-stabilized few-layered graphene nanosheets. Initially, we can tweeze graphene particles with an average size of ~2 μm in deionized (DI) water. This is accomplished using an average laser power of 300 μW, as demonstrated in Video 3. When we attempt to manipulate these same particles within a photoresist medium, the required power for successful manipulation is increased significantly. Specifically, an average power of 50 mW is needed to achieve effective control in the photoresist.

The observed variation in required OT power can be attributed to several factors that influence the optical trapping landscape. These factors include the type and size of the particles, the refractive index, and the properties of the surrounding medium. In this case, the increased viscosity of the photoresist compared to DI water likely plays a crucial role in the higher power requirement. Viscosity affects the ease with which particles can be manipulated, with higher viscosity media demanding more power to achieve similar results.

Consistent with the findings in Section 3.1, we employ the laser to trap smaller few-layered graphene particles, approximately 500 nm in size (lateral dimension of the sheets). These particles are then immobilized using STIC with an average spacing of 1 μm as seen in **Figure 4**.

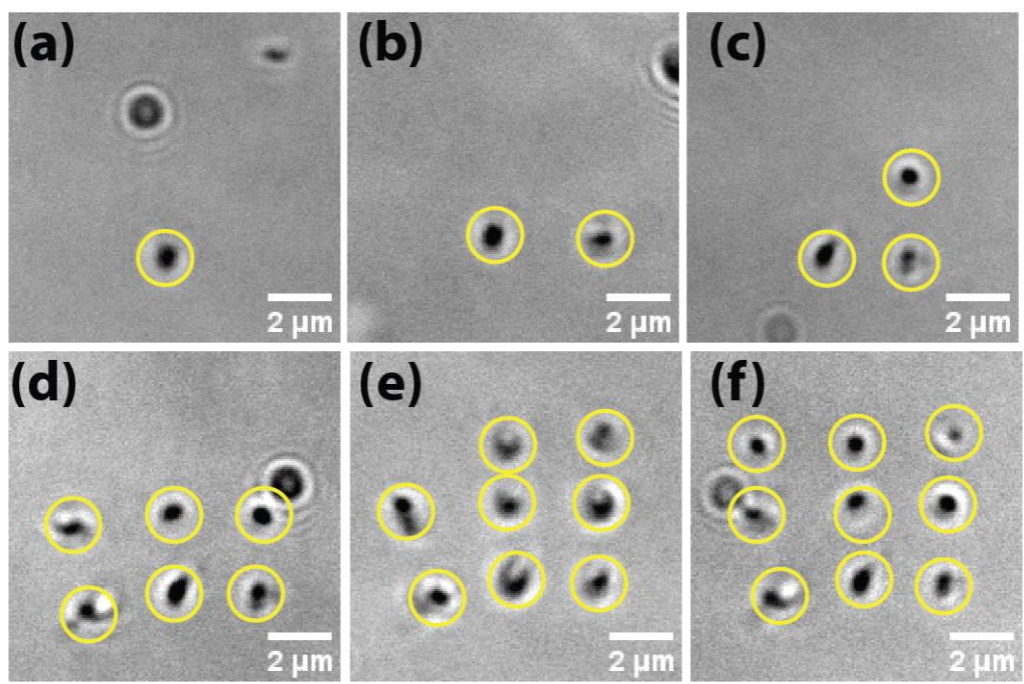

**Figure 4**. 3x3 pattern of few-layered graphene flakes.

**2.3. Systematic design of arbitrary structures with MoS$_2$**

The OT of MoS$_2$ flakes presents challenges, not only in the photoresist medium but also in deionized (DI) water. A critical factor influencing the success of OT in these scenarios is the viability of the MoS$_2$ sample. This study uses chemically exfoliated MoS$_2$ nanosheets, which are majority 1T phase and susceptible to air oxidation [18], and are thus stored under nitrogen. Once exposed to water containing dissolved O$_2$, MoS$_2$ begins to degrade by oxidation into soluble molybdate ions and sulfate, limiting the time available for optical trapping and manipulation using conventional optical tweezers due to changes in refractive index, polarizability, and structural integrities.[20] However, to avoid the complexity of introducing additional components, we adopt a different approach to manipulate $MoS_2$ flakes effectively.

Our method involves creating a "handle" using TPP, positioned 2 μm above the substrate. This handle is a crucial element in the manipulation process. Before allowing the handle to adhere to the substrate, it is manipulated using OT and brought into proximity with an $MoS_2$ flake. The flake then adheres to the handle, enabling easier manipulation of the flake (as demonstrated in Video 4). This innovative approach simplifies the manipulation process and circumvents the difficulties associated with trapping $MoS_2$ flakes directly.

Once the $MoS_2$ flakes are successfully adhered to the handle, they are immobilized in a manner similar to the procedures described in Section 3.2. The immobilization process allows for the stable positioning of the flakes, ensuring they remain in the desired configuration (**Figure 5**).

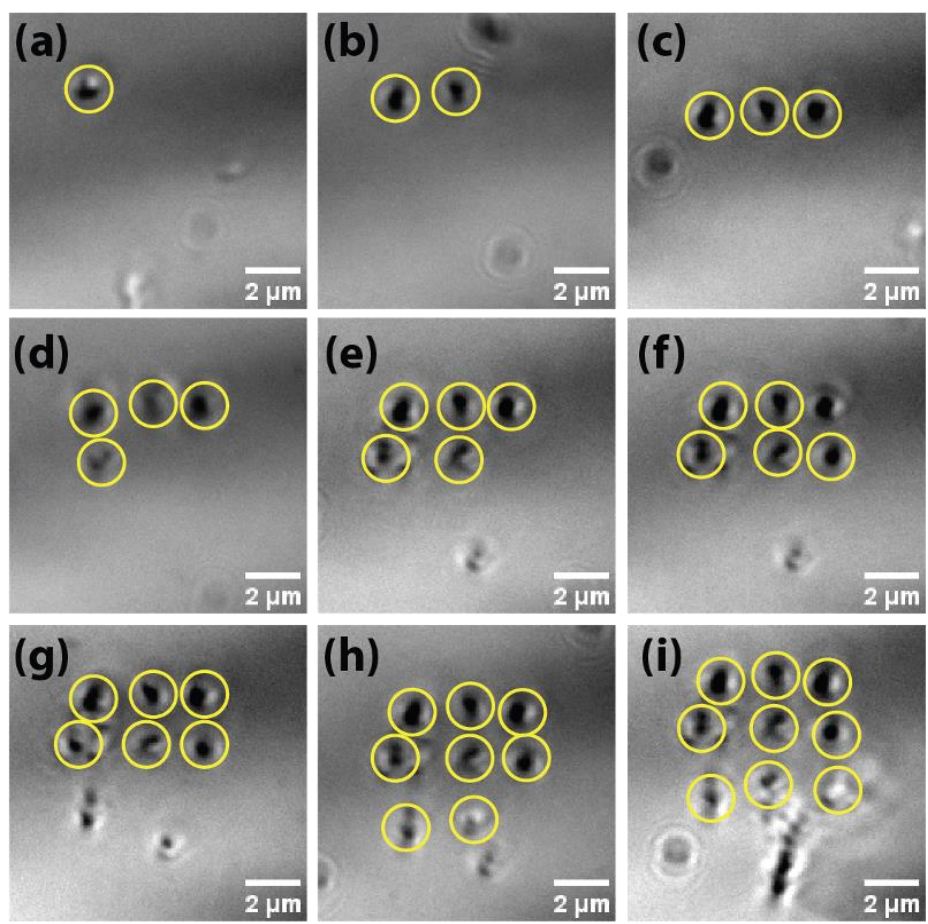

**Figure 5**. 3x3 structure of $MoS_2$ flakes of ~ 1 μm.

This technique not only overcomes the challenges posed by the oxidation of $MoS_2$ but also provides a reliable indirect approach that bypasses the difficulties associated with handling and manipulating these flakes. The simplicity and effectiveness of this technique make it a valuable tool for creating arbitrary patterns such as the one seen in Supplemental **Figure S2**.

**2.4. SHG imaging of $MoS_2$ structures**

$MoS_2$ flakes of odd layer thickness possess unique physical properties from its broken inversion symmetry, including a pronounced second-order nonlinear susceptibility.[21] The efficient frequency conversion capabilities of monolayer $MoS_2$ was leveraged by utilizing the same laser source to probe the assembled structures.

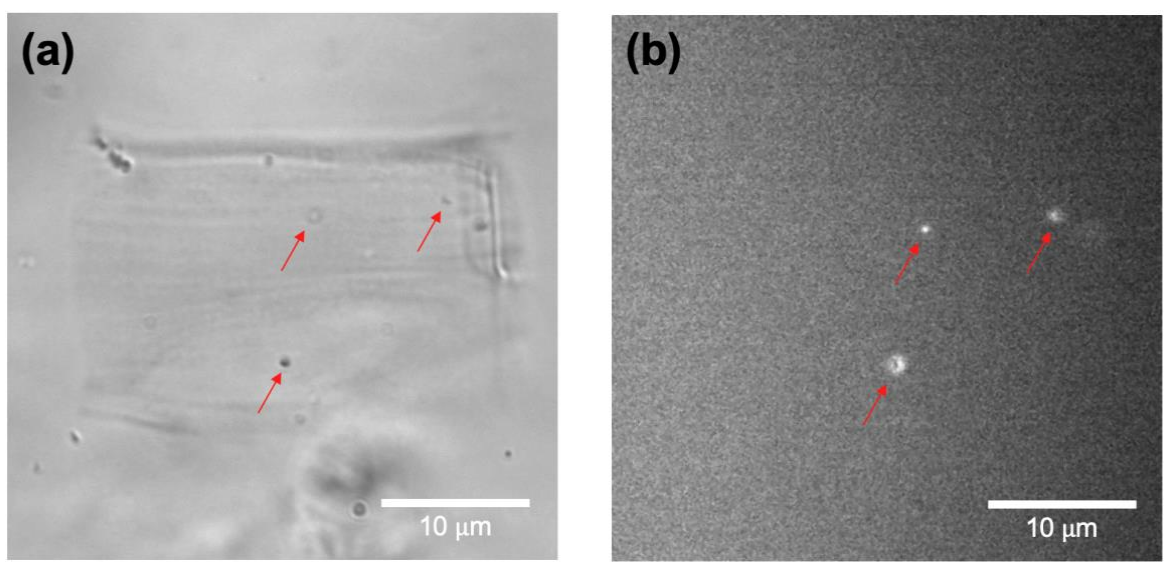

**Figure 6.** a) Brightfield image showing 3 immobilized $MoS_2$ flakes on a thin polymerized film. b) SHG image of the assembled structure. The red arrows indicate the positions of the flakes.

Three $MoS_2$ flakes are optically trapped and positioned on top of a thin polymerized film. The particles are fully immobilized by adjusting the z-position on previously printed squares of 20 μm on the flakes, as shown in **Figure 6a**. During the raster scanning process, we obtain the resultant SHG image in **Figure 6b**. The monolayer $MoS_2$ flakes exhibit strong SHG signals upon excitation.

By using a narrow bandpass filter centered at 450 nm ± 10 nm (Thorlabs, FBH450-10), the intrinsic background fluorescence emitted from the dyes in the photoresist is negligible compared to the SHG signals generated by the monolayer $MoS_2$ as seen in Figure 6b. This demonstrates the capability of this system to inspect the fabricated structures *in situ* without further post processing.

## 3. Conclusion

In summary, we developed a novel single-laser platform for the precise, contact-free assembly of 3D structures. By utilizing a femtosecond laser operating at 800-nm under ultra-low power conditions (25 mW), we successfully arranged dielectric beads into desired 2D and 3D configurations. Furthermore, the platform demonstrated its capability to handle more complex materials, including few-layer graphene and single-layer $MoS_2$ nanosheets. To overcome the challenge posed by oxidation in $MoS_2$, we introduced a TPP-fabricated handle as an intermediate holder, which effectively stabilized and positioned the $MoS_2$ flakes. We also successfully detected SHG signals from the arranged $MoS_2$ flakes using the same laser source switched to a 900-nm wavelength.

Building on its simplified setup and versatile functionality, this work lays the groundwork for an integrated micro-assembly system that minimizes heating damage and enables *in situ* inspection. Future work will focus on automating STIC using holographic optical tweezers to streamline particle manipulation, enhance reproducibility, and reduce human intervention. Additionally, by advancing the assembly of heterogeneous 3D structures that incorporate 2D materials, we aim to drive innovations in next-generation electronics and optoelectronics.

## 4. Experimental Section

*Sample Preparation*: The photoresist is composed of 8% w/v methacrylated gelatin (GelMA, Advanced BioMatrix) in phosphate buffered saline (PBS) as the prepolymer and 4.1 mM Rose Bengal (Sigma Aldrich) as the photoinitiator. Dielectric microspheres (silica 2 μm and polystyrene 5 μm) were obtained and diluted with DI water in a 1:100 ratio. Graphene oxide and $MoS_2$ nanosheets [18,19] were dispersed in DI water in a 1:1000 mass ratio. 25 μL of the photoresist and 32.5 μL sample solution were combined within a vortex mixer, and 25 μL of the final solution was drop cast onto a coverslip. The sample was covered with a gasket (Thermofisher) to prevent dehydration of the sample, and mounted with the coverslip side facing the objective lens.

*Optical Setup*: A tunable femtosecond laser operating at 120-fs pulse width and 80-MHz repetition rate (InSight X3, Spectra Physics) and at a central wavelength of 800 nm was utilized for OT and TPP. The laser is first spatially filtered with a precision pinhole of 75-μm diameter to generate a Gaussian intensity profile and expanded to a beam diameter of 3 mm. The polarization is horizontal, with respect to the laboratory optical table, using a linear polarizer, and the average power at the sample stage is controlled by the half-wave plate. The position of the laser at the sample plane is determined by a 2D galvanometer beam scanner controlled by LabVIEW. A 150 W broadband light is focused using a 20X/0.5 NA condenser lens for brightfield imaging. The backscattered light is filtered through a short pass filter (Thorlabs, FESH0700) to block the laser, and focused using a tube lens onto the sCMOS, as seen in **Figure 7**.

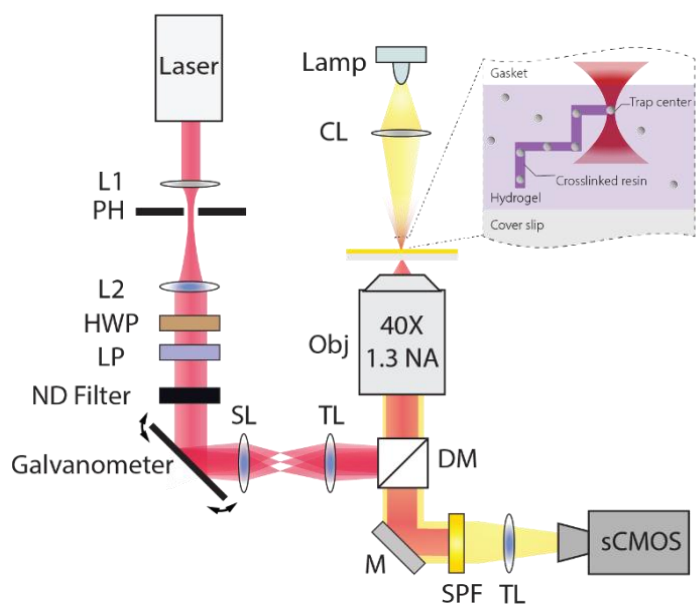

**Figure 7.** Experimental setup encompassing OT, TPP, SHG. L: lens, HWP: half wave-plate, LP: linear polarizer, ND: neutral density, SL: scan lens, TL: tube lens, DM: dichroic mirror, CL: condenser lens, M: mirror, SPF: short pass filter.

*Manufacturing Protocol:* The trapping and manipulation of particles using OT involves precise control over both the position and the average power of the laser. Initially, a particle is trapped approximately 1-2 μm above the substrate plane, with the trapping power carefully adjusted based on the type of particle being manipulated. For dielectric microspheres, an average power of 25 mW is used, while graphene particles require 50 mW, and MoS$_2$ particles require a higher power of 75 mW. This variation in power ensures that each type of particle is effectively trapped without being displaced or damaged.

Once the particle is securely trapped, it is then manipulated to the desired location within the sample. This is achieved by carefully controlling the OT movement. Simultaneously, the objective lens is moved up and down periodically. This movement induces polymerization of the hydrogel surrounding the particle, which effectively locks the particle in place by arresting its Brownian motion. The immobilization is crucial for maintaining the precise arrangement of particles, as it prevents any unwanted movement that could disrupt the structure from being fabricated.

This process is repeated for each particle involved in the fabrication until the desired structure is fully assembled. The method allows for the creation of intricate and stable microscale structures by precisely positioning and immobilizing each particle in its intended location (**Figure 8**). The ability to control particle movement and immobilization with such precision is a key advantage of this technique, enabling the fabrication of complex arrangements and structures with high accuracy.

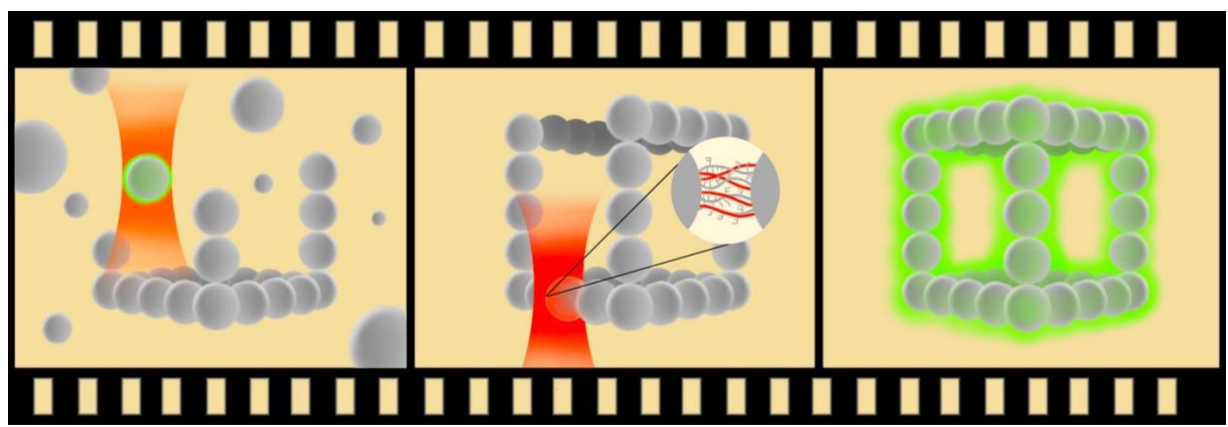

**Figure 8.** Concept illustration of the manipulation of particles to desired locations (using OT) and polymerization of the surrounding area (via TPP).

*Multiphoton microscopy inspection:* After the manufacturing of the desired structure, the laser is switched to a longer wavelength (depending on the nonlinear response of the colloids) for multiphoton microscopy inspection. The microscope shares the same optical path as the fabrication platform. A galvanometer is used to direct the beam in a raster-scan pattern. The beam is reflected by a short-pass at 800±40 nm dichroic beam splitter and focused onto the sample using an oil-immersion 1.3NA/40X objective. The emitted backward multiphoton signal is collected by the same objective. A bandpass filter is used to transmit the multiphoton signals and block the laser beam. A photomultiplier tube (Hamamatsu) records the signal and the second harmonic generation (SHG) images are subsequently reconstructed. The average power at the input of the objective is 20 mW for the fluorescent bead sample and 45 mW for the $MoS_2$ sample.

**Supporting Information**

Supporting Information is available from the Wiley Online Library or from the author.

**Author Contributions**

K.K. and J.S. designed and conducted the experiments. K.K., J.S., and W.L. analyzed the results. R.H.H provided samples. K.K., J.S., W.L., R.H.H., and K.C.T. contributed to the writing and discussion of the manuscript.

**Acknowledgements**

K.K. and J.S. contributed equally to this work. We thank Collin Polucha for careful editing of the manuscript.

**Conflict of Interest Statement**

The authors declare no conflict of interest.

**Data Availability Statement**

The data that support the findings of this study are available from the corresponding author upon reasonable request.

**References**

[1] B. Y. Majlis, in *RSM 2013 IEEE Regional Symposium on Micro and Nanoelectronics*, IEEE, **2013**, pp. vii–vii.

[2] H.-B. Yao, H.-Y. Fang, X.-H. Wang, S.-H. Yu, *Chem. Soc. Rev.* **2011**, *40*, 3764.

[3] Y. Meng, H. Zhong, Z. Xu, T. He, J. S. Kim, S. Han, S. Kim, S. Park, Y. Shen, M. Gong, Q. Xiao, S.-H. Bae, *Nanoscale Horiz* **2023**, *8*, 1345.

[4] S. Su, X. Wang, J. Xue, *Mater. Horiz.* **2021**, *8*, 1390.

[5] J. Picker, M. Schaal, Z. Gan, M. Gruenewald, C. Neumann, A. George, F. Otto, R. Forker, T. Fritz, A. Turchanin, *Nanoscale Adv.* **2023**, *6*, 92.

[6] B. Munkhbat, A. B. Yankovich, D. G. Baranov, R. Verre, E. Olsson, T. O. Shegai, *Nat. Commun.* **2020**, *11*, 4604.

[7] A. T. Barton, R. Yue, S. Anwar, H. Zhu, X. Peng, S. McDonnell, N. Lu, R. Addou, L. Colombo, M. J. Kim, R. M. Wallace, C. L. Hinkle, *Microelectron. Eng.* **2015**, *147*, 306.

[8] J. Azadmanjiri, V. K. Srivastava, P. Kumar, Z. Sofer, J. Min, J. Gong, *Appl. Mater. Today* **2020**, *19*, 100600.

[9] M. Timpel, G. Ligorio, A. Ghiami, L. Gavioli, E. Cavaliere, A. Chiappini, F. Rossi, L. Pasquali, F. Gärisch, E. J. W. List-Kratochvil, P. Nozar, A. Quaranta, R. Verucchi, M. V. Nardi, *Npj 2D Mater. Appl.* **2021**, *5*, 1.

[10] S. J. Yun, H. Ko, S. Park, B. H. Lee, N. Kim, H. P. Duong, Y. Lee, S. M. Kim, K. K. Kim, *Adv. Funct. Mater.* **2024**, DOI 10.1002/adfm.202409458.

[11] D. K. Singh, G. Gupta, *Mater. Adv.* **2022**, *3*, 6142.

[12] A. Rajan, K. Underwood, F. Mazzola, P. D. C. King, *Phys. Rev. Mater.* **2020**, *4*, DOI 10.1103/physrevmaterials.4.014003.

[13] M. Baithi, D. L. Duong, *Crystals (Basel)* **2024**, *14*, 832.

[14] S. Chizari, M. P. Lim, L. A. Shaw, S. P. Austin, J. B. Hopkins, *Small* **2020**, *16*, e2000314.

[15] B.-W. Su, X.-L. Zhang, W. Xin, H.-W. Guo, Y.-Z. Zhang, Z.-B. Liu, J.-G. Tian, *J. Mater. Chem. C Mater. Opt. Electron. Devices* **2021**, *9*, 2599.

[16] K. Krishna, J. Sun, K. C. Toussaint, in *Advanced Photonics Congress 2024*, Optica Publishing Group, Washington, D.C., **2024**, p. NoTh3B.2.

[17] K. Krishna, J. A. Burrow, Z. Jiang, W. Liu, A. Shukla, K. C. Toussaint, *J. Biomed. Opt.* **2024**, *29*, DOI 10.1117/1.jbo.29.8.086501.

[18] Y. Li, H. Yuan, A. von dem Bussche, M. Creighton, R. H. Hurt, A. B. Kane, H. Gao, *Proc. Natl. Acad. Sci. U. S. A.* **2013**, *110*, 12295.

[19] Z. Wang, A. von dem Bussche, Y. Qiu, T. M. Valentin, K. Gion, A. B. Kane, R. H. Hurt, *Environ. Sci. Technol.* **2016**, *50*, 7208.

[20] M. G. Donato CNR-IPCF, Istituto per i Processi Chimico-Fisici, V. le F. Stagno D'Alcontres 37, I.-98158, Messina, Italy. maria. donato@cnr. it onofrio. marago@cnr. it, E. Messina, A. Foti, T. J. Smart, P. H. Jones, M. A. Iatì, R. Saija, P. G. Gucciardi, O. M. Maragò, *Nanoscale* **2018**, *10*, 1245.

[21] L. Mennel, M. Paur, T. Mueller, *APL Photonics* **2019**, *4*, DOI 10.1063/1.5051965.

Supporting Information

**Harnessing ultrafast optical pulses for 3D microfabrication by selective tweezing and immobilization of colloidal particles in an integrated system**

*Krishangi Krishna, Jieliyue Sun, Robert H. Hurt, Wenyu Liu, Kimani C. Toussaint Jr.*<sup>\*</sup>

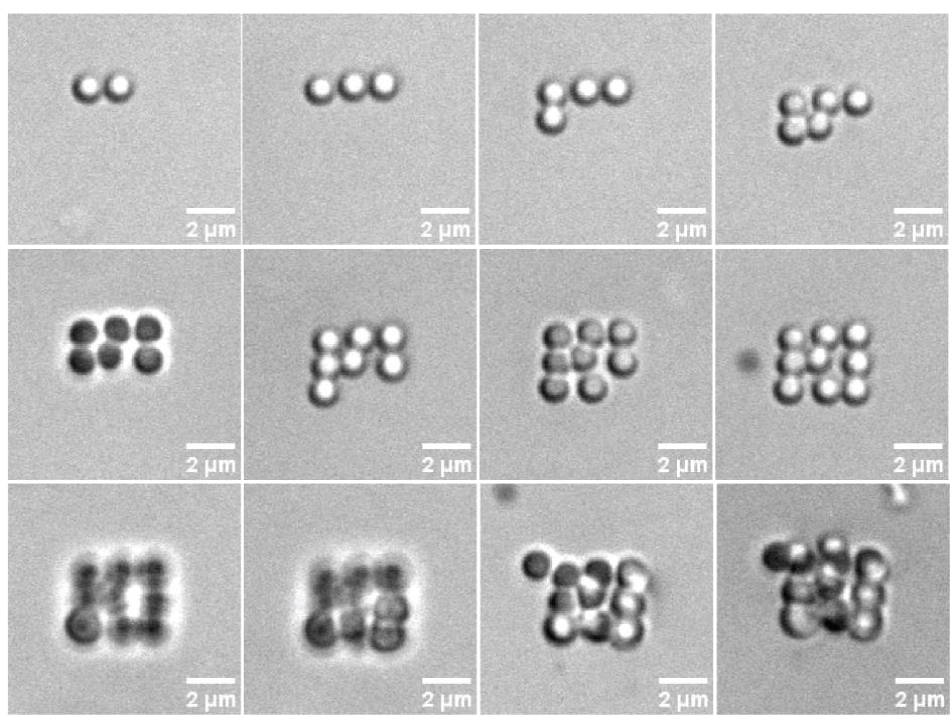

**Figure S1**. Fabrication of 3D stack of 2 μm silica spheres.

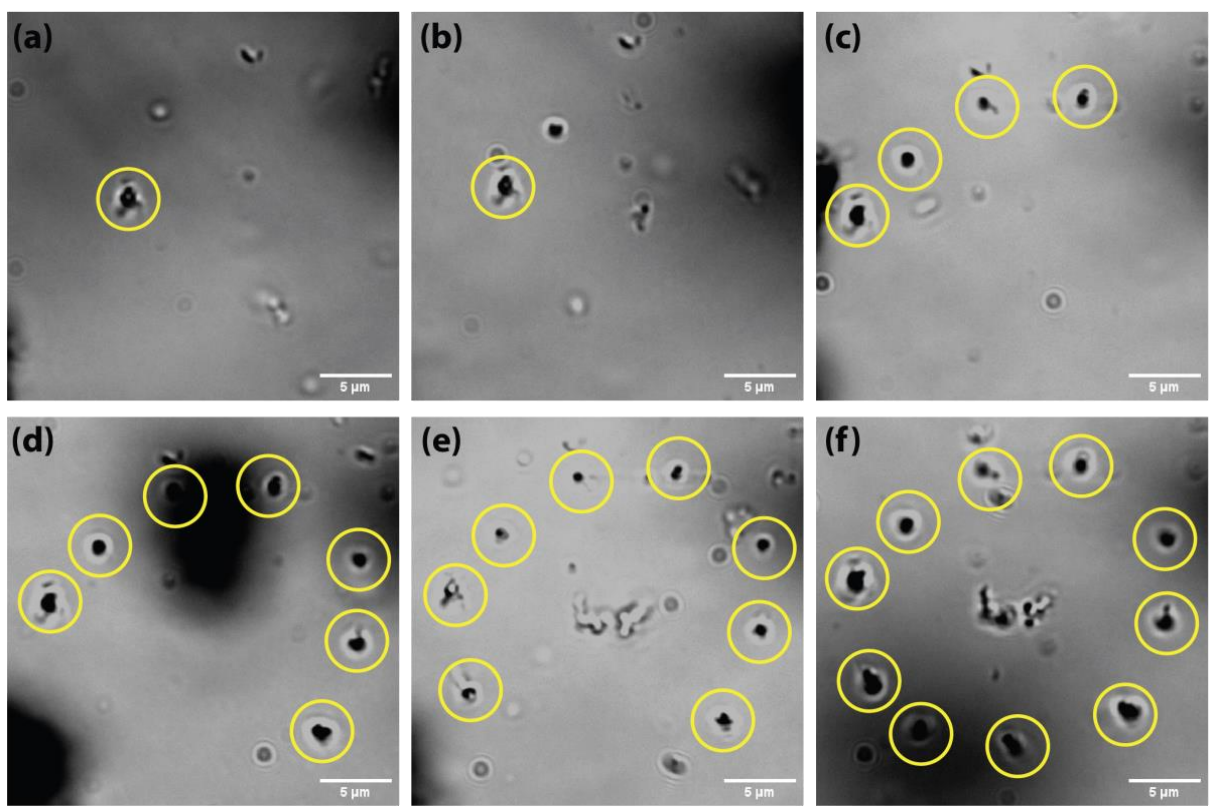

**Figure S2**. STIC can be applied on MoS$_2$ few-layered flakes to fabricate arbitrary structures.